\documentclass[final,5p,times,twocolumn]{elsarticle}
\usepackage[T1]{fontenc}
\usepackage[utf8]{inputenc}

\usepackage{amsmath}
\usepackage{amsfonts}
\usepackage{amssymb}
\usepackage{amsxtra}
\usepackage{array}
\usepackage{color}
\usepackage{dcolumn}
\usepackage{graphicx}
\usepackage{hepunits}
\usepackage{xspace}
\usepackage{CJKutf8}

\usepackage{subcaption}

\definecolor{purple}{rgb}{0.5,0,0.5}
\definecolor{blue}{rgb}{0.0,0,0.9}
\definecolor{prdblue}{rgb}{0.133,0.118,0.498}
\usepackage[colorlinks=true, pdfstartview=FitV, linkcolor=prdblue, citecolor= prdblue, urlcolor=prdblue]{hyperref}

\usepackage[mathscr,scaled=1.15]{urwchancal}
\DeclareFontFamily{OT1}{pzc}{}
\DeclareFontShape{OT1}{pzc}{m}{it}%
{<-> s * [1.15] pzcmi7t}{}
\DeclareMathAlphabet{\mathpzc}{OT1}{pzc}{m}{it}

\biboptions{sort&compress}

\journal{Physics Letters B}

\newcommand{\beq}{\begin{equation}}
\newcommand{\eeq}{\end{equation}}
\newcommand{\ba}{\begin{array}}
\newcommand{\ea}{\end{array}}
\newcommand{\bea}{\begin{align}}
\newcommand{\eea}{\end{align}}
\newcommand{\bi}{\begin{itemize}}
\newcommand{\ei}{\end{itemize}}
\newcommand{\ben}{\begin{enumerate}}
\newcommand{\een}{\end{enumerate}}
\newcommand{\bc}{\begin{center}}
\newcommand{\ec}{\end{center}}
\newcommand{\bl}{\begin{flushleft}}
\newcommand{\el}{\end{flushleft}}
\newcommand{\br}{\begin{flushright}}
\newcommand{\er}{\end{flushright}}

\begin{document}
\begin{CJK*}{UTF8}{gbsn}

\begin{frontmatter}

\title{$\,$\\[-7ex]\hspace*{\fill}{\normalsize{\sf\emph{Preprint no}. NJU-INP 112/25}}\\[1ex]
Orbital angular momentum in the pion and kaon: rest-frame and light-front}

\author[NJU,INP]{Y.-Y.\ Xiao (肖宇洋)%
       $^{\href{https://orcid.org/0009-0006-1963-7388}{\textcolor[rgb]{0.00,1.00,0.00}{\sf ID}},}$}

\author[UHe]{Z.-N.\ Xu (徐珍妮)%
    $^{\href{https://orcid.org/0000-0002-9104-9680}{\textcolor[rgb]{0.00,1,0.00}{\sf ID}},}$}

\author[HZDR]{Z.-Q.\ Yao (姚照千)%
       $\,^{\href{https://orcid.org/0000-0002-9621-6994}{\textcolor[rgb]{0.00,1,0.00}{\sf ID}},}$}

\author[NJU,INP]{C.\ D.\ Roberts%
       $^{\href{https://orcid.org/0000-0002-2937-1361}{\textcolor[rgb]{0.00,1.00,0.00}{\sf ID}},}$}

\author[UHe]{J.\ Rodr\'{\i}guez-Quintero%
       $^{\href{https://orcid.org/0000-0002-1651-5717}{\textcolor[rgb]{0.00,1,0.00}{\sf ID}},}$}

\address[NJU]{
School of Physics, \href{https://ror.org/01rxvg760}{Nanjing University}, Nanjing, Jiangsu 210093, China}
\address[INP]{
Institute for Nonperturbative Physics, \href{https://ror.org/01rxvg760}{Nanjing University}, Nanjing, Jiangsu 210093, China}

\address[UHe]{Department of Integrated Sciences and Center for Advanced Studies in Physics, Mathematics and Computation, \href{https://ror.org/03a1kt624}{University of Huelva}, E-21071 Huelva, Spain}

\address[HZDR]{\href{https://ror.org/01zy2cs03}{Helmholtz-Zentrum Dresden-Rossendorf}, Bautzner Landstra{\ss}e 400, D-01328 Dresden, Germany
\\[1ex]
\href{mailto:zhenni.xu@dci.uhu.es}{zhenni.xu@dci.uhu.es} (ZNX);
\href{mailto:z.yao@hzdr.de}{z.yao@hzdr.de} (ZQY);
\href{mailto:cdroberts@nju.edu.cn}{cdroberts@nju.edu.cn} (CDR)
\\[1ex]
Date: 2026 February 27\\[-6ex]  
}

\begin{abstract}
Orbital angular momentum (OAM) is not a Poincar\'e invariant quantity; so, its value is observer dependent.
Notwithstanding that, in quantum chromodynamics, a Poincar\'e-invariant theory, OAM is part of every hadron wave function.  Using continuum Schwinger function methods, we elucidate both the subjective character of in-hadron OAM and expose some of its impacts on pion and kaon structure and observables.  For instance, working with light-front projections of their Bethe-Salpeter wave functions, it is found that the pion is a roughly 50/50 mix of light-front OAM zero and one components and the kaon is a 60/40 system.  The overall picture is that (near) Nambu-Goldstone modes are complex bound states, each with significant intrinsic OAM, independent of the observer's reference frame.  This feature must be accounted for in the calculation of observables.  Inductively, the same is true for all hadrons.
\end{abstract}


\end{frontmatter}
\end{CJK*}

\section{Introduction}
In quark potential models, the pion, $\pi$, is represented as a $1 ^1S_0$, $J^{PC}=0^{-+}$, isospin unity ($I=1$) bound state of a constituent light quark, $l$, either up, $u$, or down, $d$, and a constituent light antiquark, $\bar l$, \textit{viz}.\
an $S$-wave, radial ground state in which the constituents possess zero orbital angular momentum (OAM), $\ell=0$, and the parity is simply given as $P=(-1)^{\ell+1}$ \cite[Ch.\,15]{ParticleDataGroup:2024cfk}.
The kaon, $K$, is similar, with the only differences being that one of the pion's light quarks or antiquarks is replaced by a strange, $s$, quark or antiquark and $I=1/2$.

In potential models, the Nambu-Goldstone (NG) boson character of the $\pi$, $K$ does not usually emerge naturally.
Instead, the low masses of these mesons \cite{ParticleDataGroup:2024cfk} are typically engineered by finely tuning pieces of the interaction potential; see, \textit{e.g}., Ref.\,\cite{Godfrey:1985xj} and citations thereof.
This need for tuning the interaction in order to reproduce NG mode masses is found in a wide variety of models, \textit{e.g}., Refs.\,\cite{Biernat:2014xaa, Choi:2014ifm, Lucha:2016vte, Ahmady:2016ufq, Lan:2019rba, dePaula:2020qna}.

Significantly and often overlooked, the NG character of these states is expressed in more than merely their low masses.
There is an intimate connection between $\pi$, $K$ properties and the axial-vector Ward-Green-Takahashi identity in quantum chromodynamics (QCD) \cite{Maris:1997hd, Maris:1997tm, Chang:2009zb, Fischer:2009jm, Qin:2014vya, Qin:2020jig, Xu:2022kng}.
This imposes many additional constraints, which extend into features of their internal structure \cite{Roberts:2021nhw, Cui:2021mom, Lu:2022cjx, Lu:2023yna, Xing:2025eip}
and also to the spectrum and structure of all their $0^{-+}$ excitations \cite{Volkov:1996br, Elias:1997ya, Holl:2004fr, Holl:2005vu, Boyle:2012jad, Ballon-Bayona:2014oma, Xu:2025cyj}.

In any Poincar\'e-invariant treatment of $\pi$, $K$ properties, these states are not purely $S$-wave systems.  That is impossible because Poincar\'e covariance itself, required for any QCD-connectable treatment of bound states seeded by light valence quarks, entails that every hadron contains OAM \cite{LlewellynSmith:1969az, Bhagwat:2006xi, Hilger:2015ora}.
Consider that if one reference frame did exist in which the OAM characterising $\pi$ and/or $K$ were zero, then because $\ell$ is not a Poincar\'e invariant quantity, one would find $\ell \neq 0$ in all other frames.  (Only total angular momentum, $J$, is observable \cite[Ch.\,1-2-2]{IZ80}.)
Furthermore, OAM and parity are unconnected in quantum field theory.
This is clear because parity is a Poincar\'e invariant quantum number, whereas $\ell$ is not; and no observable can properly be defined by a subjective quantity \cite{Brodsky:2022fqy}.

Evidently, $\pi$, $K$ structure is far more complicated than some may imagine.
For instance, the dichotomies attending upon the
 character of light/lighter quark pseudoscalar mesons are commonly obscured in effective field theories or models built upon the implicit assumption that they are elementary (pointlike) degrees of freedom.
Yet, delivering a sound explanation of the structure of strongly interacting NG bosons and its expressions in observables are fundamental longstanding challenges in high-energy nuclear and particle physics \cite{Horn:2016rip, Denisov:2018unj, Aguilar:2019teb, Anderle:2021wcy, Arrington:2021biu, Quintans:2022utc, Accardi:2023chb, Lu:2025bnm}.

Herein, we elucidate some aspects of these points by
drawing pictures of $\pi$, $K$ OAM content; identifying OAM connections with the physics of emergent hadron mass (EHM) \cite{Roberts:2020udq, Roberts:2021nhw, Binosi:2022djx, Ding:2022ows, Ferreira:2023fva, Raya:2024ejx, Achenbach:2025kfx};
calculating OAM decompositions of pseudoscalar meson electric charges and leptonic decay constants,
and explicating the dependence of these things on the quantisation/reference frame.
In addition, by considering a (fictitious) heavy pion, $\pi_{s \bar s}$, built from valence degrees of freedom with degenerate current-masses that are inflated to match that of the $s$ quark, we expose additional impacts of Higgs boson couplings into QCD on NG boson OAM content.

Interest in exposing the OAM content of Poinca\'e-covariant hadron wave functions has been high since the ``proton spin crisis'' surfaced \cite{EuropeanMuon:1987isl}.  
Using CSMs, a first elementary exposition was provided for octet and decuplet baryons in Ref.\,\cite{Oettel:1998bk}, with recent results sketched in Ref.\,\cite{Cheng:2025sdp}.
The pion and its first radial excitation were discussed in Ref.\,\cite{Bhagwat:2006xi} and, subsequently, related systems were considered in Ref.\,\cite{Hilger:2015ora}; however, these studies were restricted to the leading-order CSM approximation, \emph{viz}.\ rainbow-ladder (RL) truncation \cite{Munczek:1994zz, Bender:1996bb}.
The study herein is novel and broader because it is the first to discuss OAM decompositions obtained in a higher-level, nonperturbatively improved CSM truncation \cite{Xu:2022kng} and the first to provide internally consistent comparisons between the rest-frame and light-front pictures, with the latter enabled by the methods introduced and exploited in Ref.\,\cite{Yao:2025xjx}.

Before proceeding, we note that the results presented herein are to be interpreted as expressing meson structural properties at the hadron scale, $\zeta_{\cal H}<m_N$, $m_N$ is the nucleon mass.
At $\zeta_{\cal H}$, all hadron properties are carried by its quasiparticle valence degrees of freedom (dof).
The existence of such a scale is guaranteed by the theory of effective charges in QCD \cite{Grunberg:1980ja, Grunberg:1982fw}, \cite[Sec.\,4.3]{Deur:2023dzc}.
Evolution to scales $\zeta > \zeta_{\cal H}$ can proceed via the all-orders (AO) scheme \cite{Yin:2023dbw}, which has proven efficacious in numerous applications; see, \textit{e.g}., Refs.\,\cite{Cui:2022bxn, Xu:2023bwv, Lu:2023yna, Xu:2024nzp, Yao:2024ixu}.

\section{NG bosons via the Bethe-Salpeter equation}
\label{Sec2}
We use continuum Schwinger function methods (CSMs) to obtain Poincar\'e-covariant Bethe-Salpeter wave functions (BSWFs) for the $J^P=0^-$ states considered herein.
Each such BSWF has the form:
{\allowdisplaybreaks
\begin{subequations}
\label{BSWF}
\begin{align}
{\mathpzc X}_{\mathsf 5}& (k;Q)  =
S_f(k_\eta) \Gamma_5(k;Q) S_g(k_{\bar\eta})\\
& = S_f(k_\eta) \big\{\gamma_5 \big[
i {E}_{\mathsf 5}(k;Q) + \gamma\cdot Q {F}_{\mathsf 5}(k;Q) \nonumber \\
&
\qquad + \! \gamma\cdot k \, k\cdot Q  {G}_{\mathsf 5}(k;Q) + \sigma_{\mu\nu} k_\mu Q_\nu {H}_{\mathsf 5}(k;Q) \big]\big\}S_g(k_{\bar\eta})
\label{Eq1b}\\
& =: \gamma_5 \big[
i {\cal E}_{\mathsf 5}(k;Q) + \gamma\cdot Q {\cal F}_{\mathsf 5}(k;Q) \nonumber \\
&
\qquad +\!  \gamma\cdot k \, k\cdot Q {\cal G}_{\mathsf 5}(k;Q) + \sigma_{\mu\nu} k_\mu Q_\nu {\cal H}_{\mathsf 5}(k;Q) \big]\,, \label{XOAM} \\
&
 =: {\mathpzc g}_1 {\mathpzc E}_{\mathsf 5}(k;Q) + {\mathpzc g}_2 {\mathpzc F}_{\mathsf 5}(k;Q) \nonumber \\%
&
\qquad + {\mathpzc g}_3 {\mathpzc G}_{\mathsf 5}(k;Q) + {\mathpzc g}_4 {\mathpzc H}_{\mathsf 5}(k;Q)\,, \label{XOAM2} \\
&
=: {\mathpzc X}_{\mathsf 5}^{1}(k;Q) + {\mathpzc X}_{\mathsf 5}^{2}(k;Q) + {\mathpzc X}_{\mathsf 5}^{3}(k;Q) + {\mathpzc X}_{\mathsf 5}^{4}(k;Q) \,,\label{XOAM3}
\end{align}
\end{subequations}
where $\Gamma_5(k;Q)$ is the BS amplitude (amputated BSWF);
$Q$ is the meson total momentum, $Q^2 = -m_{\mathsf 5}^2$, $m_{\mathsf 5}$ is the meson mass;
$k$ is the relative momentum between the valence quark, $f$, and antiquark, $\bar g$, whose properties specify the character of the system;
$k_\eta = k+\eta Q$, $k_{\bar\eta} = k-(1-\eta) Q$, $0\leq \eta \leq 1$;
and ${\mathpzc g}_{i=1,2,3,4}$ are Dirac matrices, defined implicitly by comparing Eqs.\,\eqref{XOAM}, \eqref{XOAM2}.
For $m_f = m_g$, it is useful to choose $\eta=1/2$; then, $E_{\mathsf 5}, F_{\mathsf 5}, G_{\mathsf 5}, H_{\mathsf 5}$ in Eq.\,\eqref{Eq1b} are even functions of $k\cdot Q$.
}

Considering the rest-frame projection of the BSWF in Eq.\,\eqref{BSWF}, one finds \cite{Bhagwat:2006xi, Krassnigg:2009zh} that ${\cal E}_{\mathsf 5}, {\cal F}_{\mathsf 5}$ correspond to $S$-wave, whereas ${\cal G}_{\mathsf 5}, {\cal H}_{\mathsf 5}$ are $P$-wave components.
(These associations are signalled by the appearance of $k$ in the second line of Eq.\,\eqref{XOAM}.)
In contrast, regarding light-front wave functions (LFWFs), as discussed in Sect.\,\ref{Sec4} below, ${\cal F}_{\mathsf 5}$, ${\cal G}_{\mathsf 5}$ express zero OAM between the valence quasiparticle constituents and ${\cal H}_{\mathsf 5}$ generates the OAM unity component.  The remaining term, ${\cal E}_{\mathsf 5}$, does not contribute to the leading-twist LFWF.
These remarks highlight the frame dependence of OAM.
(\emph{N.B}. In light-front quantisation, with $\hat z$, the $z$-direction, taken to define the light-front, both $J$ and $J_z=\ell_z+{\mathpzc s}_z$ are kinematic operators \cite{Coester:1992cg}.
Hence, the $J=0$ pseudoscalar meson LFWF has two components:
$\ell_z=0,{\mathpzc s}=0$ and $\ell_z=\pm 1,{\mathpzc s}=\mp 1$; see Eq.\,\eqref{EqLFWF} below.
Consequently, pseudoscalar meson light-front OAM may be directly linked with the helicity of the dof.
In contrast, rest-frame OAM deals with eigenvalues of the dof $\vec{\ell}$.)

In Eq.\,\eqref{BSWF}, $S_{\{f,g\}=\{l, s\}}$ is the $2$-point Schwinger function (propagator) associated with the valence quasiparticle dof in the bound-state, with $l=u$ or $d$:
\begin{subequations}
\label{Sprop}
\begin{align}
    S_{l,s}(q)&  = 1/[i \gamma\cdot q A_{l,s}(q^2)+ B_{l,s}(q^2) ]\,,\\
    & =: Z_{l,s}(q^2)/[i \gamma\cdot q + M_{l,s}(q^2) ]\,.
\end{align}
\end{subequations}
The vector part of the dressed-quark self-energy, $A_{l,s}(q^2)$, becomes uniformly closer to unity as the quark current mass is increased, whereas the scalar part, $B_{l,s}(q^2)$, increases uniformly in magnitude at the same time  \cite[Fig.\,2.5]{Roberts:2021nhw}, \cite[Fig.\,3]{Bhagwat:2003vw}.
$M_{l,s}(q^2)$ is the dressed-quark mass function \cite{Politzer:1976tv}, which is renormalisation group invariant in a multiplicatively renormalisable theory.

Using CSMs, the solution of a given meson bound state problem is obtained by considering a set of coupled gap and Bethe-Salpeter equations \cite{Roberts:1994dr, Roberts:2012sv}.
Owing to feedup within the infinite tower of equations (Dyson-Schwinger equations, DSEs \cite{Roberts:1994dr}) of which the gap and Bethe-Salpeter equations form a part, it is necessary to introduce an approximation scheme that enables one to deliver physics results from a finite subsystem of DSEs.

Two methods are commonly used today:
(\textsf{I}) rainbow-ladder (RL) truncation, \textit{i.e}., the leading-order approximation in the scheme introduced in Refs.\,\cite{Munczek:1994zz, Bender:1996bb};
and (\textsf{II}) the nonperturbative extension (bRL) of that truncation elucidated and employed in Refs.\,\cite{Qin:2020jig, Xu:2022kng, Xu:2025cyj}.
Efficacious implementations of RL truncation can be traced from Ref.\,\cite{Maris:1997tm} and the beginning of steps beyond RL truncation are described in Refs.\,\cite{Chang:2009zb, Fischer:2009jm, Chang:2011ei}.
In both cases, two elements are key, \textit{viz}.\
(\textit{a}) the effective charge;
and
(\textit{b}) the dressed gluon-quark vertex, $\Gamma_\nu(q,p)$.

Studies of the gauge sector in QCD have led to the following efficacious form for the product of effective charge and gluon $2$-point Schwinger function \cite{Qin:2011dd, Binosi:2014aea, Binosi:2016wcx, Yao:2024ixu}:
\begin{align}
\label{defcalG}
 {\mathpzc d}(y) & =
 \frac{2\pi}{\omega^4} D e^{-y/\omega^2} + \frac{\pi \gamma_m \mathcal{F}(y)}{\tfrac{1}{2}\ln\big[ \tau+(1+y/\Lambda_{\rm QCD}^2)^2 \big]}\,,
\end{align}
where $\gamma_m=12/25$, $\Lambda_{\rm QCD} = 0.234\,$GeV, $\tau={\rm e}^2-1$, and ${\cal F}(y) = \{1 - \exp(-y/\Lambda_{\mathpzc I}^2)\}/y$, $\Lambda_{\mathpzc I}=1\,$GeV.

Given Eq.\,\eqref{defcalG}, then the kernel of the gap equation for $S_{l,s}(p)$ can be written as $(l=(p-q), y=l^2)$ \cite{Maris:1997tm, Binosi:2016wcx}:
\begin{equation}
\label{DSEkernel}
{\mathpzc d}(y) T_{\mu\nu}(l) [i\gamma_\mu\frac{\lambda^{a}}{2} ]_{tr} [i\Gamma_\nu(q,p)\frac{\lambda^{a}}{2} ]_{su}\,,
\end{equation}
$l^2 T_{\mu\nu}(l) = l^2 \delta_{\mu\nu} - l_\mu l_\nu$.
Landau gauge is employed because, amongst other merits, it is a fixed point of the QCD renormalisation group.
In solving all relevant DSEs, we use a mass-independent momentum-subtraction renormalisation scheme \cite{Chang:2008ec}, with renormalisation scale $\zeta=19\,$GeV.  At this scale, the quark wave function renormalisation constant is practically unity, $Z_2\approx 1$, which leads to useful simplifications.  Evolution to different renormalisation scales is straightforward.

Rainbow-ladder truncation may now be introduced as \cite[Sec.\,II]{Maris:1997tm} $\Gamma_\nu^{\rm RL}(q,p) = \gamma_\nu$;
then the bRL (EHM-improved) vertex employed in Refs.\,\cite{Xu:2022kng, Xu:2025cyj} has the following form:
\begin{equation}
\label{EqbRL}
\Gamma_\nu(q,p) =
\Gamma_\nu^{\rm RL}(q,p)
- {\mathpzc a} \, \kappa(y) \, \sigma_{\alpha\nu} \, l_\alpha \,.
\end{equation}
Here, ${\mathpzc a} = 1.1$ expresses the strength of the associated EHM-induced dressed-quark anomalous chromomagnetic moment (ACM) \cite{Singh:1985sg, Bicudo:1998qb, Chang:2010hb, Qin:2013mta}.
The ACM is essentially nonperturbative in origin; hence, an IR-focused profile function is appropriate, \textit{viz}.\
$\kappa(y) = (1/\omega) \exp(-y/\omega^2)$.
This ACM improvement eliminates most defects of RL truncation in spectrum calculations \cite{Chang:2011ei, Xu:2022kng} and also has a discernible effect on hadron structure \cite{Xu:2025cyj}.

It should be stressed here that whilst RL truncation is still the most widely used, owing to its simplicity, available comparisons with inferences from data and other robust calculations indicate that bRL results are more realistic; see, \textit{e.g}., Refs.\,\cite{Cui:2020tdf, Cui:2021mom, Cui:2022bxn, Xu:2022kng, Xing:2023wuk, Roberts:2023lap, Xu:2025cyj}.

\begin{table}[t]
\caption{ \label{params}
{\sf Panel A}.\
Quark current masses and interaction parameters, Eq.\,\eqref{defcalG}, employed to calculate meson BSWFs.
The strength of the bRL anomalous chromomagnetic moment, Eq.\,\eqref{EqbRL}, is ${\mathpzc a} = 1.1$.
{\sf Panel B}.\
Meson masses and leptonic decay constants calculated using the values in Panel A.
With the listed renormalisation group invariant current masses, leading-order evolution to $\zeta = 2\,$GeV yields the following values (in GeV):
RL -- $0.0054, 0.121$; bRL -- $0.0031, 0.083$.
For context, Ref.\,\cite[PDG]{ParticleDataGroup:2024cfk} reports: $0.0035$, $0.094$, values which align well with the bRL masses.
%
%
(All tabulated quantities listed in GeV.)
}
\begin{tabular*}
{\hsize}
{
l@{\extracolsep{0ptplus1fil}}
|l@{\extracolsep{0ptplus1fil}}
l@{\extracolsep{0ptplus1fil}}
l@{\extracolsep{0ptplus1fil}}
l@{\extracolsep{0ptplus1fil}}}\hline\hline
\centering
{\sf A} & $\hat m$ & $\hat m_s$ & $\omega$ & $(\omega D)^{1/3}$ \\ \hline
RL  & $0.0072$ & $0.161$ & $0.6$ & $ 0.80$\\
bRL $\ $ & $0.0041$ & $0.110$ & $0.8$ & $ 0.72$\\
\hline
\end{tabular*}

\medskip

\begin{tabular*}
{\hsize}
{
l@{\extracolsep{0ptplus1fil}}
|l@{\extracolsep{0ptplus1fil}}
l@{\extracolsep{0ptplus1fil}}
l@{\extracolsep{0ptplus1fil}}
l@{\extracolsep{0ptplus1fil}}
l@{\extracolsep{0ptplus1fil}}
l@{\extracolsep{0ptplus1fil}}}\hline
\centering
{\sf B} & $m_\pi$ & $m_K$ & $m_{\pi_{s\bar s}}$
        & $f_\pi$ & $f_K$ & $f_{\pi_{s\bar s}}$ \\ \hline
RL  & $0.14$ & $0.49$ & $0.69$ & $0.090$ & $0.11 $ & $ 0.13\phantom{00}$\\
bRL $\ $ & $0.14$ & $0.49$ & $0.67$ & $0.098$ & $0.11$ & $ 0.12$\\
\hline\hline
\end{tabular*}
\end{table}

In the calculation of any observable, one must employ the canonically normalised BSWF \cite[Sec.\,3]{Nakanishi:1969ph}.
In nonrelativistic quantum mechanics, a bound-state wave function, $\Psi(x)$, is a probability amplitude and its normalisation is simply implemented by introducing a multiplicative scaling factor that ensures $1= \int dx\,|\Psi(x)|^2$.
In Poincar\'e-invariant quantum field theory, however, owing, \textit{inter alia}, to the loss of particle number conservation, the Poincar\'e-covariant Bethe-Salpeter wave function does not permit interpretation as a probability amplitude.
One consequence of this is that its normalisation is different from that used in nonrelativistic quantum mechanics.


Using any approximation scheme in which the Bethe-Salpeter kernel, ${\cal K}$, is independent of the total momentum of the pair of valence dof, $Q$, like RL truncation, it is common to implement canonical normalisation by rescaling the solution of the homogeneous Bethe-Salpeter equation such that \cite[Eq.\,(27)]{Maris:1997tm}:
\begin{align}
\label{norm-m}
    1 & =
    2 {\rm tr}_{\rm CD}
    \int_{dk} \!
    \frac{d}{dQ^2} \bigg[ i\bar{\mathpzc X}_{\mathsf 5}(k;-K)
    S_f^{-1}(k_\eta) i{\mathpzc X}_{\mathsf 5}(k;K)S_g^{-1}(k_{\bar \eta})
    \bigg]_{K=Q}^{Q^2+m_{\mathsf 5}^2=0},
%
\end{align}
where the trace is over colour and spinor indices,
$\int_{dk}$ indicates a translationally invariant regularisation of the four-dimensional integral,
and $\bar{\mathpzc X}_5(k; -Q) = C^\dagger {\mathpzc X}_5^{\rm T}(-k; -Q) C$, with $C=\gamma_2 \gamma_4$ and $(\cdot)^T$ indicating matrix transpose,
is the charge-conjugated BSWF \cite[Eq.\,(27)]{Maris:1997tm}.

In the general case, when ${\cal K}$ does depend on $Q$, canonical normalisation can be implemented by first writing the homogeneous Bethe-Salpeter equation in the following eigenvalue form:
\begin{equation}
\Gamma_{\mathsf 5}(k;Q)
= \lambda(Q^2) \int_{dq} {\mathpzc X}_{\mathsf 5} (q;Q)\, {\cal K}(q,k;Q)\,.
\end{equation}
Then the required meson wave function is obtained at that value of $Q^2$ for which $\lambda(Q^2)=1$ and the resulting wave function is normalised by rescaling such that \cite[Sec.\,3]{Nakanishi:1969ph}:
\begin{align}
\label{eq:Nakanishi Normalization}
1 & = \bigg[\frac{d \ln\lambda(Q^2)}{d Q^2}
\nonumber \\
& \times \textrm{tr}_\textrm{CD}\int_{dk}
\bar{\mathpzc X}_{\mathsf 5}(k;-Q)
S_f^{-1}(k_\eta) {\mathpzc X}_{\mathsf 5}(k;Q)S_g^{-1}(k_{\bar \eta})
\bigg]_{Q^2+m_{\mathsf 5}^2=0}\,.
%
%
%
\end{align}
The analogue for baryons can be found in Ref.\,\cite[Fig.\,3]{Wang:2018kto}.

Importantly, in RL truncation and using the same Bethe-Salpeter wave function, both Eqs.\,\eqref{norm-m} and \eqref{eq:Nakanishi Normalization} yield an identical numerical result for the normalisation rescaling factor.  This is despite the fact that, after the differentiation in Eq.\,\eqref{norm-m} is applied, which is common when proving current conservation in RL truncation \cite{Roberts:1994hh}, the integrands in Eqs.\,\eqref{norm-m} and \eqref{eq:Nakanishi Normalization} are very different.
One merit of the procedure associated with Eq.\,\eqref{eq:Nakanishi Normalization} is that it has more of the appearance of wave function normalisation in quantum mechanics.

In addition to RL truncation, herein we also deliver predictions obtained with the bRL kernel defined implicitly by Eq.\,\eqref{EqbRL} and constructed explicitly using the procedure introduced in Refs.\,\cite{Qin:2020jig, Xu:2022kng}.
In each channel, this bRL kernel introduces additional sensitivity to both EHM and Higgs-boson related chiral symmetry breaking effects, but its dependence on the meson total momentum is weak.
Consequently,
the numerical value of the canonical normalisation constant obtained using the general formula, Eq.\,\eqref{eq:Nakanishi Normalization}, is little different from that yielded by the $Q$-independent kernel form, Eq.\,\eqref{norm-m}: for all systems considered herein, the relative differences are $\lesssim 1$\%.
The mismatch is especially small because all three systems are pseudoscalar meson ground states and the ACM has most observable impact on systems in which EHM plays a greater role, like axial-vector mesons \cite{Chang:2011ei, Qin:2020jig, Xu:2022kng}.

At this point, with the quark current masses and parameter values in Table~\ref{params}\,A, the necessary gap and Bethe-Salpeter equations can be solved using standard algorithms \cite{Maris:1997tm, Maris:2005tt, Krassnigg:2009gd}.
Along with the pointwise behaviour of the meson Bethe-Salpeter wave functions, this delivers the meson masses (from the value of $Q^2$ for which $\lambda(Q^2)=1$) and leptonic decay constants listed in Table~\ref{params}\,B.
The leptonic decay constants are obtained using (recall that $Z_2\approx 1$):
\begin{equation}
\label{Eqf5}
f_{\mathsf 5} Q_\mu
= {\rm tr}_{\rm CD}\int_{dk} \gamma_5\gamma_\mu X_{\mathsf 5}(k;Q)\,.
\end{equation}

For the ground state $\pi$, $K$, both RL and bRL predictions are a good match with empirical determinations \cite{ParticleDataGroup:2024cfk}.  This outcome is an implicit expression of the NG character of these states \cite{Maris:1997hd, Maris:1997tm, Chang:2009zb, Fischer:2009jm, Qin:2014vya}: all symmetry-preserving CSM truncations must deliver a good description.
Regarding $\pi_{s\bar s}$, the meson masses are the same, but one sees a $12$\% difference between the decay constants.
This is one indication of the observable impact of bRL corrections in the neighbourhood of $\hat m_s$, \textit{i.e}.\ following this size of Higgs-boson induced shift away from the NG boson limit.

In order to calculate charge radii, one must know the electromagnetic current that is symmetry-consistent with the approximation to the Bethe-Salpeter kernel.  
For RL truncation, this current was introduced in Ref.\,\cite{Roberts:1994hh}: it may be viewed as the origin of the Eq.\,\eqref{norm-m} normalisation condition; and has since been widely and efficaciously employed, most recently in Ref.\,\cite{Yao:2024drm}.  
The symmetry-consistent current is not known for any bRL kernel.  
Notwithstanding, given the near numerical equivalence of the Eq.\,\eqref{norm-m} and Eq.\,\eqref{eq:Nakanishi Normalization} normalisation values obtained using the bRL kernel described herein, one might reasonably expect that the associated bRL electromagnetic current evaluated with the associated wave functions yields results that are quantitatively similar to those obtained in RL truncation.  
Here, therefore, we record charge radii of the $\pi$, $K$, $\pi_{s\bar s}$, which were calculated elsewhere \cite{Chen:2018rwz, Yao:2024drm} using precisely the same RL framework as that described herein (in fm):
$r_\pi = 0.67$; $r_K = 0.60$; $r_{\pi_{s\bar s}}=0.49$.
Where the comparison makes sense, these results are in good agreement with the empirical values \cite{ParticleDataGroup:2024cfk, Cui:2022fyr}: $r_\pi = 0.663(6)$; $r_K = 0.56(3)$.

\section{Rest-frame OAM}
\label{Sec3}
Consider the following collection of projection operators:
\begin{subequations}
    \begin{align}
    {\mathcal P}_{1={\cal E} = S_{\rm wave}} & = \tfrac{-i}{4}\gamma_5  \,,\\
    {\mathcal P}_{2={\cal F}=S_{\rm wave}} & = \tfrac{1}{4Q^2}\gamma\cdot Q \gamma_5 \,, \\
    {\mathcal P}_{3={\cal G} = P_{wave}} & = \tfrac{1}{4}\tfrac{1}{k^2}\tfrac{1}{k\cdot Q} \gamma\cdot k \gamma_5 \,,\\
    {\mathcal P}_{4={\cal H} = P_{\rm wave}} & =\tfrac{1}{4}\tfrac{1}{(k\cdot Q)^2-k^2 Q^2} \sigma_{\mu\nu} Q_\mu k_\nu \,,
    \end{align}
\end{subequations}
which, when applied to Eq.\,\eqref{XOAM}, yield the independent wave function components whose rest-frame OAM association is specified in the subscript.
Working with these projection operators, one has
\begin{equation}
{\mathpzc X}_{\mathsf 5}^i (k;Q) = {\mathpzc g}_i {\rm tr}_{\rm D}{\mathcal P}_{i}{\mathpzc X}_5(k;Q)\,,
\end{equation}
with no sum on $i$, \textit{e.g}., the leading $S$-wave component is
${\mathpzc X}_{\mathsf 5}^1 (k;Q) = {\mathpzc g}_1 {\mathpzc E}_5(k;Q)$.

\begin{figure}[t]
    \centering
    \begin{subfigure}{0.49\columnwidth}
        \includegraphics[width=\textwidth]{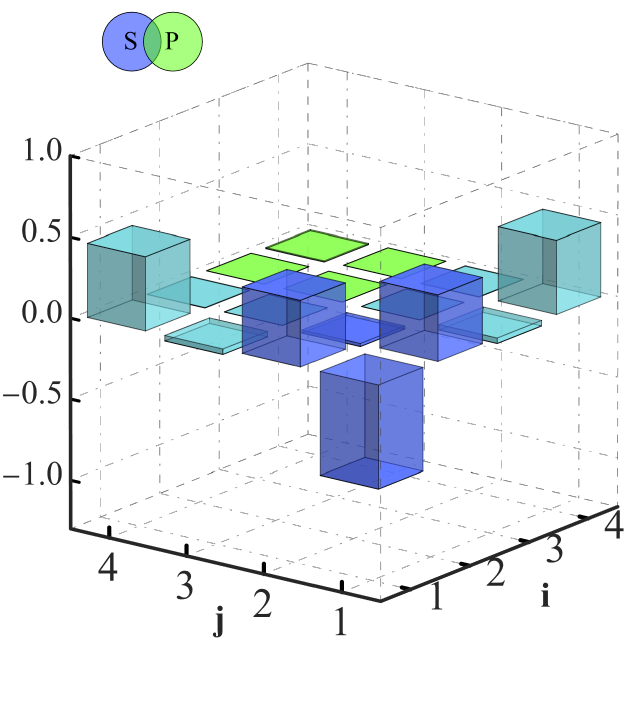}
        \caption{\small $\pi^+$}
        \label{fig3:subim1}
    \end{subfigure}
    \begin{subfigure}{0.49\columnwidth}
        \includegraphics[width=\textwidth]{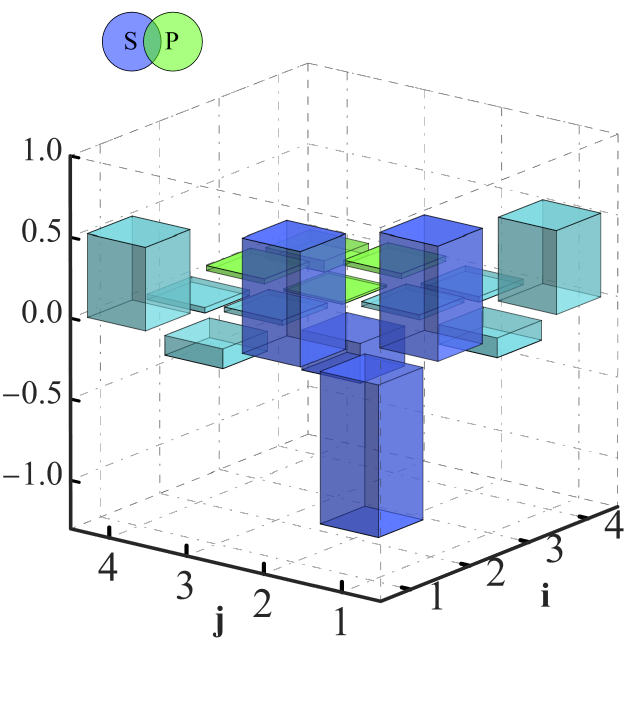}
        \caption{\small $K^+$}
        \label{fig3:subim2}
    \end{subfigure}
    \begin{subfigure}{0.49\columnwidth}
        \includegraphics[width=\textwidth]{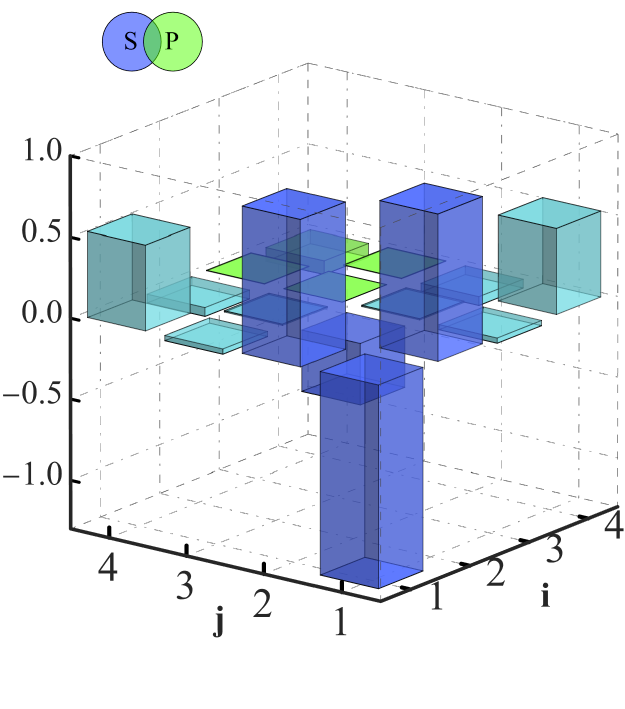}
        \caption{\small $\pi_{s\bar{s}}^+$}
        \label{fig3:subim3}
    \end{subfigure}
    \caption{RL truncation valence dof OAM pairing contributions to the normalisation of a charged pseudoscalar meson, defined via Eq.\,\eqref{LIJRL}.
    There are both positive (above plane) and negative (below plane) contributions, the total sum of which is unity.
\label{figomega6}}
\end{figure}

Focusing first on RL truncation, then using these projection operators and the Poincar\'e-covariant Bethe-Salpeter wave function, we construct the following matrix:
\begin{align}
    & {\mathpzc l}^{ij}=  2 {\rm tr}_{\rm CD} \int_{dk} \! \frac{d}{dQ^2}
    \nonumber \\
    &
    \qquad \times \left[
i\bar{\mathpzc X}_{\mathsf 5}^i(k;-Q)
S_f^{-1}(k_\eta) i{\mathpzc X}_{\mathsf 5}^j(k;Q)S_g^{-1}(k_{\bar \eta})
\right]_{Q=P}^{P^2+m_{\mathsf 5}^2=0}\,.
\label{DefLIJO}
\end{align}
Equation~\eqref{norm-m} guarantees
\begin{equation}
    1 = \sum_{i,j}^4 {\mathpzc l}^{ij}\,.
    \label{LIJRL}
\end{equation}
Hence, when calculated with RL truncation Bethe-Salpeter wave functions evaluated in the meson rest frame, ${\mathpzc l}^{ij}$ may be viewed as a way of measuring the contribution of a given OAM partial-wave pairing to the bound-state normalisation.  Since canonical normalisation is equivalent to requiring that the elastic electric form factor of a charged pseudoscalar meson is unity at zero momentum transfer, then, equivalently, ${\mathpzc l}^{ij}$ provides one means of describing the contribution of a given OAM partial-wave pairing to the bound-state electric charge.
In implementing this scheme, we employ a modified version of the procedure introduced in Ref.\,\cite{Hilger:2015ora}.

\begin{table}[t]
\caption{ \label{C1OAM}
{\sf Panel A}.
Contributions (measured in \%) from various in-meson rest-frame partial wave overlaps to the pseudoscalar meson charge (canonical normalisation) as measured in RL truncation by the matrix in Eq.\,\eqref{DefLIJO}.  The entries quantify the images in Fig.\,\ref{figomega6}.
{\sf Panel B}. \% contributions from in-meson rest-frame partial wave overlaps to the meson charge as measured in RL truncation by the matrix in Eq.\,\eqref{DefLIJ} -- quantifying Fig.\,\ref{figomega6A} images.
{\sf Panel C}. \% contributions from in-meson rest-frame partial wave overlaps to the meson charge as measured in bRL truncation by the matrix in Eq.\,\eqref{DefLIJ} -- quantifying Fig.\,\ref{figomega6B} images.
}
\begin{tabular*}
{\hsize}
{
l@{\extracolsep{0ptplus1fil}}
|c@{\extracolsep{0ptplus1fil}}
c@{\extracolsep{0ptplus1fil}}
c@{\extracolsep{0ptplus1fil}}}\hline\hline
\centering
{\sf A}\, RL\, Eq.\,\eqref{DefLIJO} $\ $ & $S \otimes S\ $ & $S \otimes P\ $ & $P \otimes P\ $ \\\hline
$\pi^+$ & $16.2\ $ & $84.5\ $ & $-0.7 \ $ \\
$K^+$ & $23.7\ $ & $78.3\ $ & $-2.0 \ $ \\
$\pi_{s\bar s}^+$ & $17.9\ $ & $90.1\ $ & $-8.0 \ $ \\
\hline\hline
\end{tabular*}
\medskip

\begin{tabular*}
{\hsize}
{
l@{\extracolsep{0ptplus1fil}}
|c@{\extracolsep{0ptplus1fil}}
c@{\extracolsep{0ptplus1fil}}
c@{\extracolsep{0ptplus1fil}}}\hline\hline
\centering
{\sf B}\, RL\, Eq.\,\eqref{DefLIJ} $\ $ & $S \otimes S\ $ & $S \otimes P\ $ & $P \otimes P\ $ \\\hline
$\pi^+$ & $100.7\ $ & $-1.9\ $ & $\phantom{1}1.2 \ $ \\
$K^+$ & $\phantom{1}97.5\ $ & $-0.3\ $ & $\phantom{1}2.8 \ $ \\
$\pi_{s\bar s}^+$ & $108.9\ $ & $-22.1\phantom{1}\ $ & $13.2 \ $ \\
\hline\hline
\end{tabular*}
\medskip

\begin{tabular*}
{\hsize}
{
l@{\extracolsep{0ptplus1fil}}
|c@{\extracolsep{0ptplus1fil}}
c@{\extracolsep{0ptplus1fil}}
c@{\extracolsep{0ptplus1fil}}}\hline\hline
\centering
{\sf C}\, bRL\, Eq.\,\eqref{DefLIJ} $\ $ & $S \otimes S\ $ & $S \otimes P\ $ & $P \otimes P\ $ \\\hline
$\pi^+$ & $101.0\ $ & $-1.6\ $ & $\phantom{1}0.6 \ $ \\
$K^+$ & $110.3\ $ & $-17.4\phantom{1}\ $ & $\phantom{1}7.1 \ $ \\
$\pi_{s\bar s}^+$ & $119.8\ $ & $-33.5\phantom{1}\ $ & $13.7 \ $ \\
\hline\hline
\end{tabular*}
\end{table}

Figure~\ref{figomega6}\,A, depicts RL truncation results for the valence quasiparticle dof OAM-pairing contributions to the normalisation of the $\pi^+$, defined via Eqs.\,\eqref{DefLIJO}, \eqref{LIJRL}.
As with analogous images for baryons \cite{Cheng:2025sdp}, there are both positive (above plane) and negative (below plane) contributions, and the total sum of all contributions is unity (100\%).
Evidently, the ${\mathpzc X}_\pi^1 \otimes {\mathpzc X}_\pi^1$ ($S \otimes S$-wave) contribution has the largest magnitude but it is also negative.
This part is marginally overcompensated by positive ${\mathpzc X}_\pi^1 \otimes {\mathpzc X}_\pi^2$ contributions.
So, finally, the unit normalisation (charge) is almost entirely provided by
${\mathpzc X}_\pi^1 \otimes {\mathpzc X}_\pi^4$ ($S \otimes P$-wave) constructive interference; see Table~\ref{C1OAM}\,A.

As valence quark current mass is increased -- see Figs.\,\ref{figomega6}\,B, C, the magnitude of both
${\mathpzc X}^1 \otimes {\mathpzc X}^1$ and ${\mathpzc X}^1 \otimes {\mathpzc X}^2$ terms increases and an ${\mathpzc X}^2 \otimes {\mathpzc X}^2$ term appears.
The sum of these four contributions remains a modestly sized positive number; so that the positive normalisation (charge) continues to be largely generated  by
${\mathpzc X}^1 \otimes {\mathpzc X}^4$ ($S \otimes P$-wave) constructive interference.
(This terms grows slower with increasing current mass.)
In all cases, $P \otimes P$-wave interference is destructive when measured via Eq.\,\eqref{DefLIJO}.

Note now that:
(\textit{i}) the axial-vector Ward-Green-Takahashi identity forces an approximate linear proportionality between ${\mathpzc X}^1$ and the scalar piece of the dressed-quark self-energy, Eq.\,\eqref{Sprop} -- see, \textit{e.g}., Ref.\,\cite[Eq.\,(34)]{Maris:1997tm} -- and the latter function increases with quark current mass;
and
(\textit{ii}) the ${\mathpzc X}^2$ term carries a linear dependence on $m_{\sf 5}$.
Consequently, it is unsurprising that these $S\otimes S$ terms increase in magnitude in a manner that approximately matches the growth in current mass.
(Naturally, this is hidden in the relative sizes (percentages) listed in Table~\ref{C1OAM}.)

Owing to charge conjugation symmetry in $\pi, \pi_{s\bar s}$, $\mathpzc X^3$-connected contributions to the normalisation are only apparent in the kaon.  These lead to destructive interference, which suppresses $S\otimes P$-wave strength, so a greater fraction of the charge owes to the $S\otimes S$-wave components.

Now, instead, consider the general case, Eq.\,\eqref{eq:Nakanishi Normalization}, working from which one can construct an analogous matrix:
\begin{align}
    & {\mathpzc L}^{ij}
    = - \left[ \frac{d \ln\lambda(Q^2)}{d Q^2} \right. \nonumber \\
&    \left. \times \, \textrm{tr}_\textrm{CD}\int_{dk}
\bar{\mathpzc X}_{\mathsf 5}^i(k;-Q)
S_f^{-1}(k_\eta) {\mathpzc X}_{\mathsf 5}^j(k;Q)S_g^{-1}(k_{\bar \eta})
\right]_{P^2+m_{\mathsf 5}^2=0}\,,
\label{DefLIJ}
\end{align}
wherewith, this time, Eq.\,\eqref{eq:Nakanishi Normalization} guarantees
    $1 = \sum_{i,j}^4 {\mathpzc L}^{ij}\,.$

\begin{figure}[t]
    \centering
    \begin{subfigure}{0.49\columnwidth}
        \includegraphics[width=\textwidth]{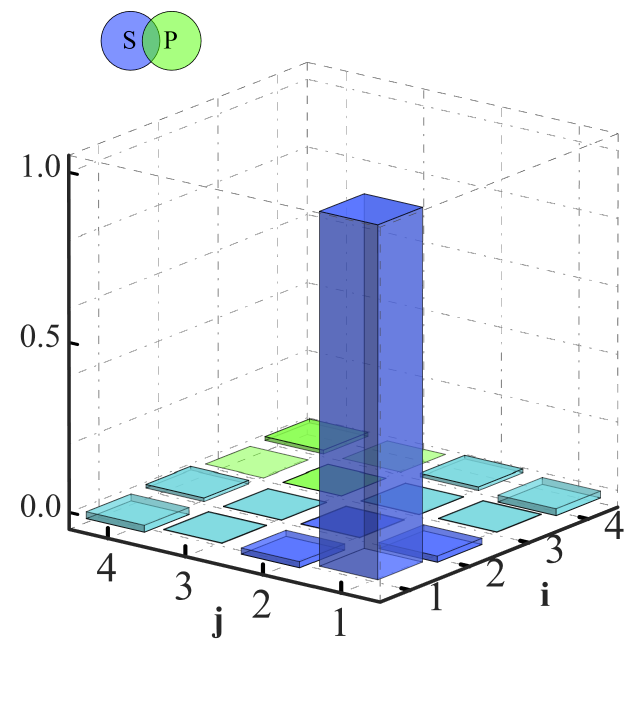}
        \caption{\small $\pi^+$}
        \label{fig3:subim1}
    \end{subfigure}
    \begin{subfigure}{0.49\columnwidth}
        \includegraphics[width=\textwidth]{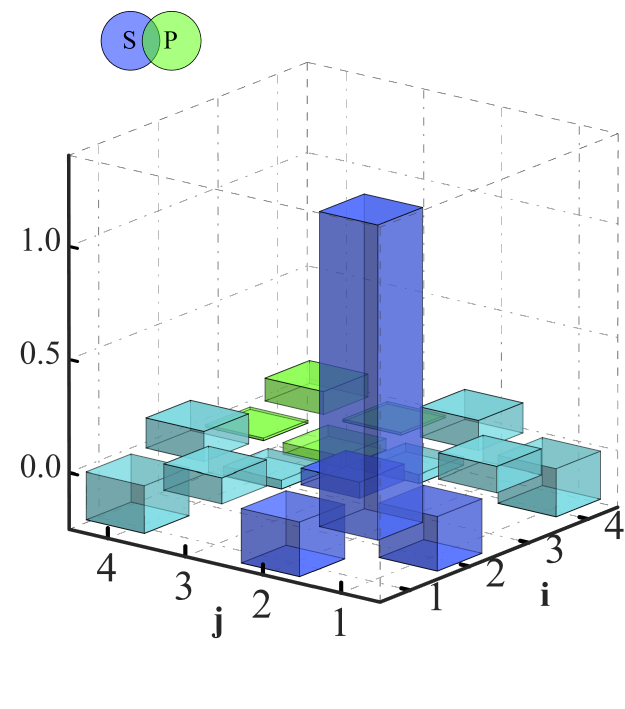}
        \caption{\small $K^+$}
        \label{fig3:subim2}
    \end{subfigure}
    \begin{subfigure}{0.49\columnwidth}
        \includegraphics[width=\textwidth]{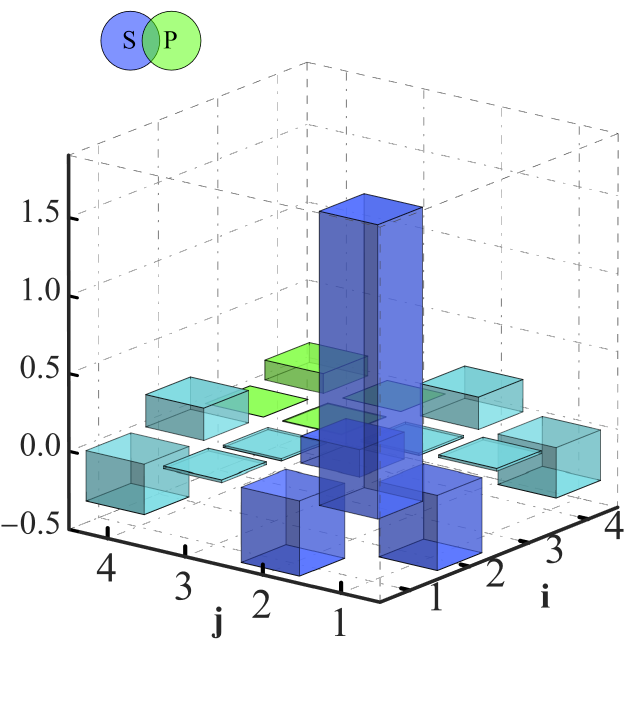}
        \caption{\small $\pi_{s\bar{s}}^+$}
        \label{fig3:subim3}
    \end{subfigure}
    \caption{RL truncation valence dof OAM pairing contributions to the normalisation of a charged pseudoscalar meson, defined via Eq.\,\eqref{DefLIJ}.
    As in Fig.\,\ref{figomega6}, there are both positive (above plane) and negative (below plane) contributions -- see Table~\ref{C1OAM}, the total sum of which is unity, notwithstanding the marked differences between related images.
\label{figomega6A}}
\end{figure}

\begin{figure}[t]
    \centering
    \begin{subfigure}{0.49\columnwidth}
        \includegraphics[width=\textwidth]{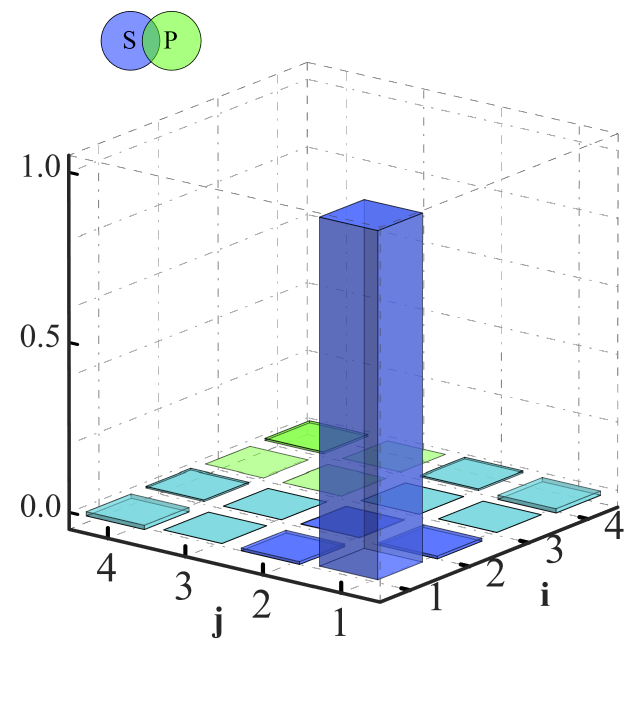}
        \caption{\small $\pi^+$}
        \label{fig3:subim1}
    \end{subfigure}
    \begin{subfigure}{0.49\columnwidth}
        \includegraphics[width=\textwidth]{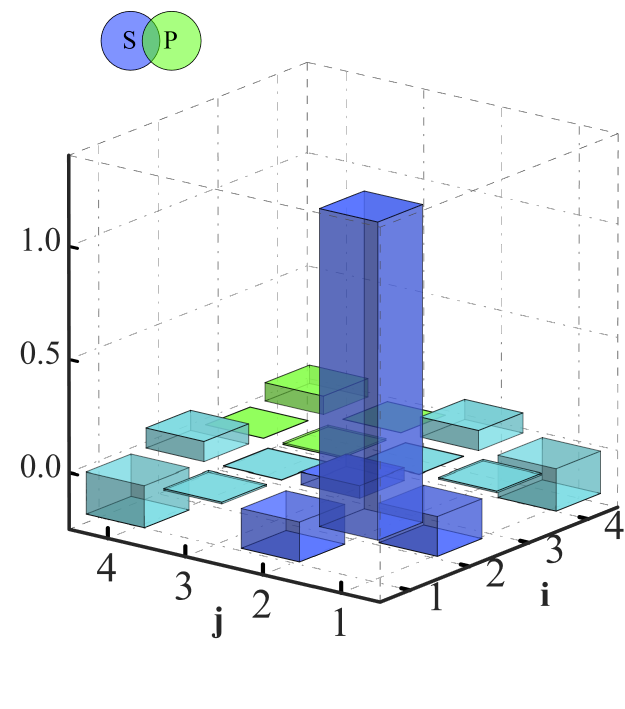}
        \caption{\small $K^+$}
        \label{fig3:subim2}
    \end{subfigure}
    \begin{subfigure}{0.49\columnwidth}
        \includegraphics[width=\textwidth]{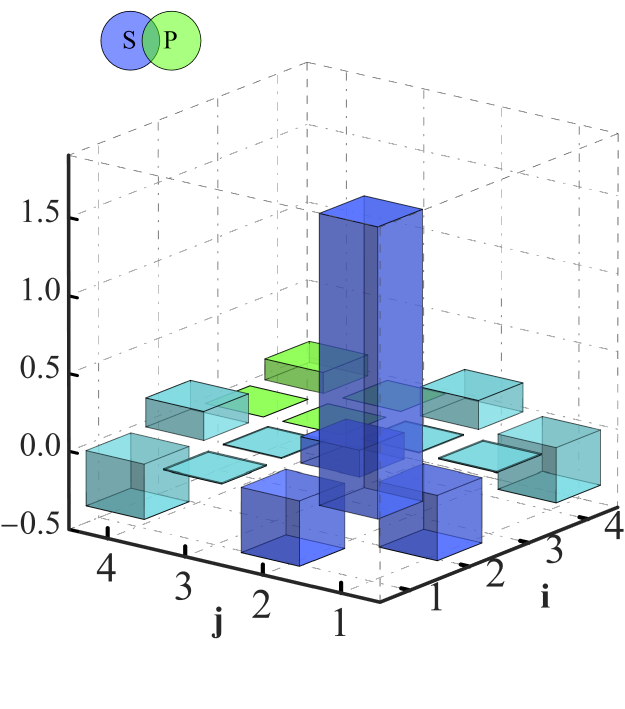}
        \caption{\small $\pi_{s\bar{s}}^+$}
        \label{fig3:subim3}
    \end{subfigure}
    \caption{bRL truncation valence dof OAM pairing contributions to the normalisation of a charged pseudoscalar meson, defined via Eq.\,\eqref{DefLIJ}.
    As in Fig.\,\ref{figomega6}, there are both positive (above plane) and negative (below plane) contributions -- see Table~\ref{C1OAM}, the total sum of which is unity, notwithstanding the marked differences between related images.
\label{figomega6B}}
\end{figure}

The matrix in Eq.\,\eqref{DefLIJ} provides an alternative measure of the rest-frame OAM pairing contributions to a meson's electric charge.
In omitting the derivative on the integrand when compared with the matrix in Eq.\,\eqref{DefLIJO}, it is plain that Eq.\,\eqref{DefLIJ} is more akin to the normalisation condition of bound-state wave functions in nonrelativistic quantum mechanics.

Calculated RL results obtained using Eq.\,\eqref{DefLIJ} are drawn in the panels of Fig.\,\ref{figomega6A}; see, also, Table~\ref{C1OAM}\,B.
By this measure, the rest-frame pion is predominantly $S\otimes S$-wave, although $P$-waves contribute somewhat.
Here, too, with increasing dof current mass, the relative strengths of $P$-wave contributions are magnified.
However, the images in Fig.\,\ref{figomega6A} are very different from those in Fig.\,\ref{figomega6}.
This highlights again that, in quantum field theory, where canonical normalisation involves combining numerous integrals of overlaps of the various Bethe-Salpeter wave function components, it is only the sum that is truly a probability, not the individual terms in that sum.

Some additional remarks are, perhaps, useful at this point.
The integrands in Eqs.\,\eqref{norm-m}, \eqref{DefLIJO} are obtained by considering the electromagnetic currents of the valence constituents inside the meson \cite{Mandelstam:1955sd}.
However, as stressed in Ref.\,\cite[Sec.\,3]{Nakanishi:1969ph}, there are many other possibilities, \emph{e.g}.,  avoiding issues with electric-charge-zero bound-states, the energy-momentum tensor was used in Ref.\,\cite{NAMBU1964214}.
These two schemes, and others, too, deliver different integrands, but identical values for the integral therewith defined \cite{PredazziBSE}.  
Likewise, for Eqs.\,\eqref{eq:Nakanishi Normalization}, \eqref{DefLIJ}.
(Such outcomes are standard; for instance, $\int_0^1 2x = 1=\int_0^1[4x - 3x^2]$.)
This must be the case; otherwise, the normalisation of a Bethe-Salpeter amplitude would be arbitrary and, \emph{e.g}., one could not use it to predict even a leptonic decay constant.
The basic point is that a system's OAM decomposition depends on both the frame used to define it and the probe used to resolve it. 
Thus, OAM decompositions in Poincar\'e-invariant quantum field theory are only defined within these subjective horizons.

Returning to Eqs.\,\eqref{eq:Nakanishi Normalization}, \eqref{DefLIJ}, OAM decompositions obtained using the bRL kernel are drawn in Fig.\,\ref{figomega6B}.
Evidently, for the pion, by this measure, RL \textit{cf}.\ bRL OAM decompositions are qualitatively similar.  Nevertheless, there are quantitative differences -- compare Tables~\ref{C1OAM}\,B, C: as seen in Ref.\,\cite{Yao:2025xjx}, the bRL kernel enhances OAM effects in the bound state because EHM remains the dominant matter-mass generating mechanism on a larger domain of current mass than in RL truncation.

\begin{figure}[t]
    \centering
    \begin{subfigure}{0.495\columnwidth}
        \includegraphics[width=\textwidth]{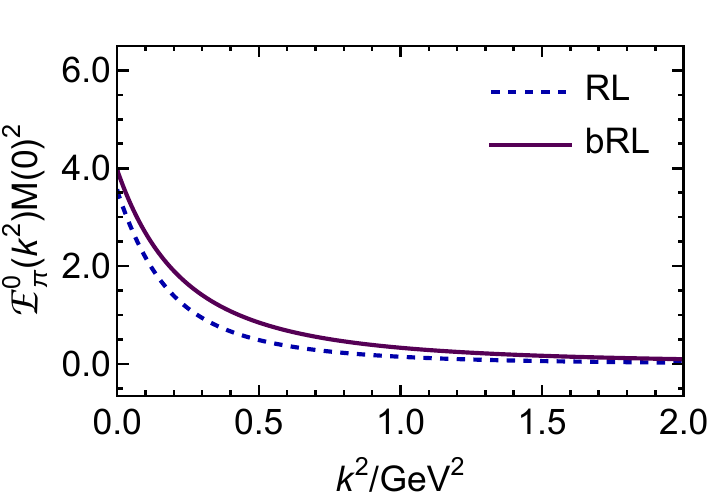}
        \caption{\small $\pi^+$}
        \label{fig3:subim1}
    \end{subfigure}
    \begin{subfigure}{0.495\columnwidth}
        \includegraphics[width=\textwidth]{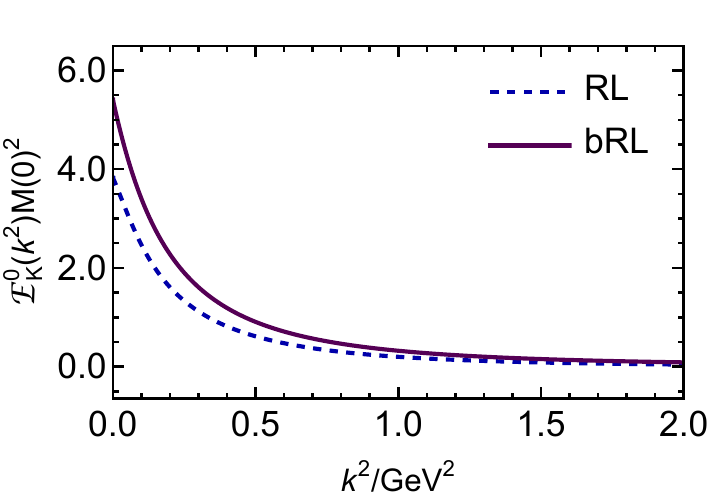}
        \caption{\small $K^+$}
        \label{fig3:subim2}
    \end{subfigure}
    \begin{subfigure}{0.495\columnwidth}
        \includegraphics[width=\textwidth]{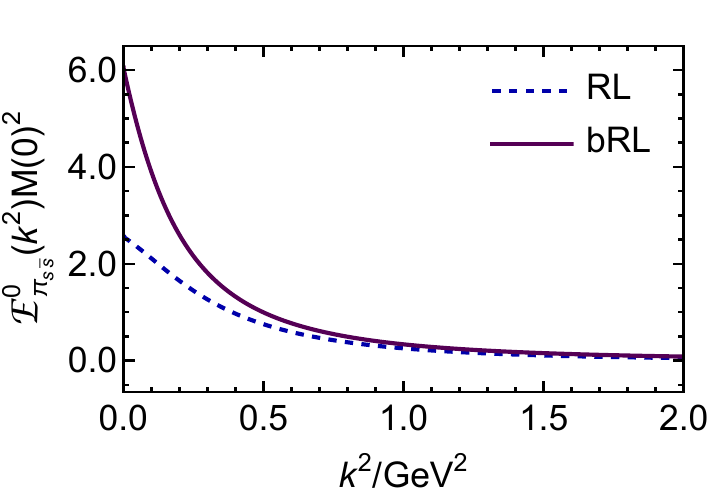}
        \caption{\small $\pi_{s\bar{s}}^+$}
        \label{fig3:subim3}
    \end{subfigure}
    \caption{Zeroth Chebyshev moment of the ${\mathpzc E}_{\mathsf 5}(k;Q)$ term in Eq.\,\eqref{XOAM2} as calculated with RL and bRL kernels.
    Using the RL kernel, $M(0)=0.4\,$GeV, and with the bRL interaction, $M(0)=0.45\,$GeV.
\label{PlotXH}}
\end{figure}

It is also worth indicating the impact of the different truncations on the wave functions themselves.  To that end, in Fig.\,\ref{PlotXH}, we depict the zeroth Chebyshev moment \cite[Eq.\,(69)]{Maris:1997tm} of the ${\mathpzc E}_{\mathsf 5}(k;P)$ term in Eq.\,\eqref{XOAM2}, denoted, as usual, ${\mathpzc E}_{\mathsf 5}^0(k;P)$.
This provides the largest magnitude contribution in all cases; see Figs.\,\ref{figomega6}\,--\,\ref{figomega6B}.
The evident differences between RL and bRL results are typical of the impacts of EHM-improvement on the wave functions as a whole.
Namely, given the axial-vector Ward-Green-Takahashi identity constraints and owing to the pion being a near NG boson, both RL and bRL truncations deliver much the same results for this system.
The impacts of EHM-improvement are larger in the current mass asymmetric kaon and they grow, too, with increasing quark current mass, in large part because the heavier systems are farther from the NG boson limit.
Given that the ACM is an infrared effect, all differences vanish with increasing $k^2$.

Another observable of interest is the leptonic decay constant, Eq.\,\eqref{Eqf5}, which can also be decomposed over rest-frame partial wave contributions.
Owing to the spinor trace, only $i=2$ (${\cal F}_{\mathsf 5}$, $S$-wave) and $i=3$ (${\cal G}_{\mathsf 5}$, $P$-wave) contribute.
The results are listed in Table~\ref{leptonicdecayX}.
Evidently, calculated using RL truncation, the $S$-wave contributions are large and positive, with the excess being cancelled via destructive interference with the $P$-wave contribution.
Although the $\pi$ and $K$ results are the same, the $S$- and $P$-wave fractional contributions diminish in magnitude as the current mass of both valence constituents is increased.

Using the bRL Bethe-Salpeter kernel, on the other hand, one sees in Table~\ref{leptonicdecayX} that the rest-frame $S$-wave contribution is a more important fraction of the final result for $f_{\mathsf 5}$, \textit{viz}.\ $S$-wave alone yields a result closer to the final value.
Furthermore, the partitioning between $S$- and $P$ wave components is practically the same for all cases considered herein.
These observations indicate that EHM expressions in the bRL kernel dominate over Higgs-boson mass generating effects on a much greater domain of quark current masses than one finds with the RL kernel.
Since bRL is a more realistic kernel, then these outcomes should be considered as more reliable indications of the impacts of rest-frame OAM components within pseudoscalar mesons.

\section{Light-front OAM}
\label{Sec4}
As explicated, \textit{e.g}., in Ref.\,\cite{Yao:2025xjx}, with a Poincar\'e-covariant Bethe-Salpeter wave function in hand, it is possible to obtain the bound-state's LFWF via a straightforward numerical procedure.  For a pseudoscalar meson, the resulting leading-twist valence quasiparticle LFWF has the form:
\begin{align}
\psi_{\mathsf 5}(x,k_\perp^2)
& \propto
\gamma_5 \big[\gamma\cdot n \,\psi_{\mathsf 5}^{\mathpzc L = 0}(x,k_\perp^2)\nonumber \\
& \qquad + i \sigma_{\mu\nu}n_\mu k_{\perp \nu} \, \psi_{\mathsf 5}^{\mathpzc L = 1}(x,k_\perp^2)\big]\,,
\label{EqLFWF}
\end{align}
where $n$ is a lightlike $4$-vector, $n^2=0$, $n\cdot k_\perp = 0$, and $n\cdot P=-m_{\mathsf 5}$ in the meson rest frame.
The superscript in Eq.\,\eqref{EqLFWF} indicates the light-front orbital angular momentum projection:
${\mathpzc L}=0=\uparrow\downarrow=\downarrow\uparrow$ means the LFWF has the light-front spins of the valence constituents antialigned;
and ${\mathpzc L}=1=\uparrow\uparrow=\downarrow\downarrow$ has them aligned.

We take this opportunity to again highlight the frame-dependence of OAM.  Regarding Eq.\,\eqref{EqLFWF}, ${\mathpzc X}_5^2$ (rest-frame $S$-wave) and ${\mathpzc X}_5^3$ (rest-frame $P$-wave) contribute to the light-front OAM ${\mathpzc L}=0$ term and ${\mathpzc X}_5^4$ (rest-frame $P$-wave) is the sole contributor to light-front ${\mathpzc L}=1$.
Since the ${\mathpzc X}_5^1$ term is higher twist \cite{Chang:2013epa}, it makes no contribution to $\psi_{\mathsf 5}(x,k_\perp^2)$.

\begin{table}[t]
\caption{ \label{leptonicdecayX}
Rest-frame OAM decomposition of meson leptonic decay constants, listed as a multiplicative fraction of the total result in Table~\ref{params}\,B.
}
\begin{tabular*}
{\hsize}
{
l@{\extracolsep{0ptplus1fil}}
|c@{\extracolsep{0ptplus1fil}}
c@{\extracolsep{0ptplus1fil}}
c@{\extracolsep{0ptplus1fil}}
c@{\extracolsep{0ptplus1fil}}}\hline\hline
\centering
   &       & $f_\pi$ & $f_K$ & $f_{\pi_{s\bar s}}$  \\ \hline
RL$\ $ & $S\ $ & $\phantom{-}1.33\ $ & $\phantom{-}1.34\ $& $\phantom{-}1.25\ $ \\
   & $P\ $ & $-0.33$ & $-0.34\ $ & $-0.25\ $ \\
\hline
bRL$\ $ & $S\ $ & $\phantom{-}1.16\ $ & $\phantom{-}1.18\ $& $\phantom{-}1.18\ $ \\
   & $P\ $ & $-0.16$ & $-0.18\ $ & $-0.18\ $ \\
\hline\hline
\end{tabular*}
\end{table}

\begin{table}[t]
\caption{ \label{LFWFOAM}
OAM decomposition of meson LFWF normalisation, Eq.\,\eqref{DefTMD}.
}
\begin{tabular*}
{\hsize}
{
l@{\extracolsep{0ptplus1fil}}
|c@{\extracolsep{0ptplus1fil}}
c@{\extracolsep{0ptplus1fil}}
c@{\extracolsep{0ptplus1fil}}
c@{\extracolsep{0ptplus1fil}}
c@{\extracolsep{0ptplus1fil}}
c@{\extracolsep{0ptplus1fil}}}\hline\hline
\centering
   & \multicolumn{2}{c}{$\pi$}
   & \multicolumn{2}{c}{$K$}
   & \multicolumn{2}{c}{$\pi_{s\bar s}$} \\
   & $P_0$ & $P_1$ & $P_0$ & $P_1$ & $P_0$ & $P_1$ \\\hline
$\ $ RL$\ $ & $0.64\ $ & $0.36\ $ & $0.66\ $& $0.34\ $ & $0.73\ $ & $0.27\ $ \\
bRL    & $0.51\ $ & $0.49\ $ & $0.57\ $ & $0.43\ $ & $0.62\ $ & $0.38\ $\\
\hline\hline
\end{tabular*}
\end{table}

In comparison with the BSWF, a bound-state's LFWF does have a probability interpretation.
Consequently, its normalisation is guaranteed by an identity like that in nonrelativistic quantum mechanics:
\begin{align}
    1 & = \frac{1}{(2\pi)^3}\int dx \int d^2 k_\perp
    \left[ | \psi_{\mathsf 5}^{0}(x,k_\perp^2)|^2
+ k_\perp^2 | \psi_{\mathsf 5}^{1}(x,k_\perp^2)|^2\right] \,.
\label{DefTMD}
\end{align}
In this case, it is straightforward to identify the relative strength of the different light-front OAM components.  We list the results in Table~\ref{LFWFOAM}, obtained by using the procedure in Ref.\,\cite{Yao:2025xjx} to project the Bethe-Salpeter wave functions onto the light-front.

Three important observations are signalled by Table~\ref{LFWFOAM}.
(\textit{i}) Since Eq\,\eqref{DefTMD} is equivalent to ensuring unit electric charge for pseudoscalar mesons, then ${\mathpzc L}=1$ components of all states considered herein are crucial parts of their LFWFs.  They may be ignored only with foresight and careful compensation, and exploiting the results in Ref.\,\cite{Yao:2025xjx}.
(\textit{ii}) The relative strength of the ${\mathpzc L}=1$ light-front OAM component diminishes with increasing quark current mass and the ${\mathpzc L}=0$ contribution grows to compensate.
(\textit{iii}) The EHM-improved bRL Bethe-Salpeter kernel leads to a material enhancement of ${\mathpzc L}=1$ components in pseudoscalar meson LFWFs.  Indeed, the ground-state pion is roughly an equal mixture of $\mathpzc L=0$ and $\mathpzc L=1$ light-front OAM components.
This OAM picture is markedly different from that which one draws from Fig.\,\ref{figomega6B}.

Regarding $f_{\mathsf 5}$ in Eq.\,\eqref{Eqf5}, the light-front picture is also simpler than the rest-frame decomposition.  One has
\begin{equation}
f_{\mathsf 5} = \int dx \int_{d k_\perp} {\rm tr}_{\rm D} \tfrac{1}{5}
\gamma\cdot n \gamma_5 \psi_{\mathsf 5} (x,k_\perp^2)\,;
\end{equation}
namely, instead of results like those in Table~\ref{leptonicdecayX}, the light-front ${\mathpzc L}=0$,  $\psi_{\mathsf 5}^{0}$, piece of the wave function produces the entire value of $f_{\mathsf 5}$.
Thus, when using LFWFs, there is a stark contrast between the roles of OAM in producing electric and weak charges of pseudoscalar mesons.
Notwithstanding these things, as noted above, the $\psi_{\mathsf 5}^{0}$ piece of the LFWF is just the object obtained from the rest-frame BSWF $\ell=0,1$ components, $\mathpzc X_{\mathsf 5}^2$, $\mathpzc X_{\mathsf 5}^3$, which contribute to the decay constant defined by Eq.\,\eqref{Eqf5}.
This is why the value of $f_{\mathsf 5}$ is unchanged, even though its interpretation in terms of frame-dependent OAM contributions is different.

\section{Perspective}
The ``proton spin crisis'' \cite{EuropeanMuon:1987isl} has made the orbital angular momentum content of hadron bound states into a much debated topic.
In these discussions, a variety of issues are often overlooked; so, we reiterate a few here.
(\textit{a}) Orbital angular momentum (OAM) is not a Poincar\'e invariant quantity; so, it is frame (hence, observer) dependent.
(\textit{b}) In QCD, a relativistic quantum field theory, like the degrees of freedom amongst which it is distributed, OAM depends on the energy scale of the probe used to infer its frame-dependent value.

Herein, using continuum Schwinger function methods (CSMs), we focused on (\textit{a}), elucidating aspects of the subjective character of in-hadron OAM by exposing its structural impacts within Nature's most fundamental (near) Nambu-Goldstone (NG) bosons, \textit{viz}. pions and kaons.  (In the CSM context, elements of (\textit{b}) are sketched elsewhere \cite{Yu:2024ovn}.)

The analysis revealed and stressed the following points.
Today, using CSMs, it is straightforward to calculate a meson's Poincar\'e covariant Bethe-Salpeter wave function (BSWF) [Sect.\,\ref{Sec2}].
Analysing such BSWFs, it becomes clear that structural features of NG bosons are qualitatively insensitive to the kernel used to complete the CSM bound-state equations [Sect.\,\ref{Sec3}].
Naturally, there are quantitative differences, but they are modest because (near) NG boson properties are much constrained by identities associated with partial conservation of the axial-vector current.
Notwithstanding these facts, the perceived OAM decomposition within a given (near) NG mode is very sensitive to the approach chosen to define it [Fig.\,\ref{figomega6} \textit{cf}.\ Figs.\,\ref{figomega6A}, \ref{figomega6B}].
Consequently, whilst it is straightforward to calculate a BSWF, that wave function cannot readily be interpreted because it is not a probability amplitude.

A system's LFWF is significantly harder to calculate, but its interpretation is clean because LFWFs do admit interpretation as probability amplitudes, albeit, typically, in the infinite-momentum frame.
Fortunately, using modern CSMs and with a BSWF in hand, it is possible to calculate the associated LFWF for the system of interest.
Using this approach, the pion is seen to be a roughly 50/50 mix of light-front OAM zero and one components and the kaon is a 60/40 system.

Our analysis supports a picture of (near) NG modes as complex systems, each with a significant amount of intrinsic OAM, independent of the observer's reference frame.  That OAM must be properly understood and accounted for in the calculation of observables.  The same is indubitably true for all hadrons.

\medskip

\noindent\textbf{Acknowledgments}.
We thank K.\ Raya for constructive comments.
Work supported by:
National Natural Science Foundation of China, grant no.\ 12135007;
Helmholtz-Zentrum Dresden-Rossendorf, under the High Potential
Programme;
and
Ministerio Espa\~nol de Ciencia, Innovaci\'on y Universidades (MICINN) grant no.\ PID2022-140440NB-C22.

\medskip
\noindent\textbf{Data Availability Statement}. This manuscript has no associated data or the data will not be deposited. [Authors' comment: All information necessary to reproduce the results described herein is contained in the material presented above.]

\medskip
\noindent\textbf{Declaration of Competing Interest}.
The authors declare that they have no known competing financial interests or personal relationships that could have appeared to influence the work reported in this paper.


\end{document}